\documentstyle[11pt,paspconf,epsf]{article}
\input{psfig}

\newcommand{\gtsim}{\ {\raise-0.5ex\hbox{$\buildrel>\over\sim$}}\ }
\newcommand{\ltsim}{\ {\raise-0.5ex\hbox{$\buildrel<\over\sim$}}\ }
\newcommand{\frog}{{\sc {frog}}}
\newcommand{\frogs}{{\sc {frog}}s}
\newcommand{\eros}{{\sc {ero}}s}

\def\simlt{\lower.5ex\hbox{$\; \buildrel < \over \sim \;$}}
\def\simgt{\lower.5ex\hbox{$\; \buildrel > \over \sim \;$}}

\begin{document}

\title{Faint Infrared-Excess Field Galaxies: \frogs}

\author{L. A. Moustakas, M. Davis, S. E. Zepf\altaffilmark{1}, \&
A. J. Bunker}
\affil{Astronomy Department, University of California, Berkeley, CA 94720\\
{\tt email: leonidas@astro.berkeley.edu}}
\altaffiltext{1}{Present address: Astronomy Department, Yale University}

\begin{abstract}
Deep near-infrared and optical imaging surveys in the field reveal a
curious population of galaxies that are infrared-bright
($I-K\gtsim4$), yet with relatively blue optical colors
($V-I\ltsim2$).  Their surface density, several per square arcminute
at $K>20$, is high enough that if placed at $z>1$ as our models
suggest, their space densities are about one-tenth of $\phi_*$.  The
colors of these ``faint red outlier galaxies'' (\frogs) may derive
from exceedingly old underlying stellar populations, a dust-embedded
starburst or AGN, or a combination thereof.  Determining the nature of
these \frogs\ has implications for our understanding of the processes
that give rise to infrared-excess galaxies in general.  We report on
an ongoing study of several targets with HST \& Keck imaging and
Keck/LRIS multislit spectroscopy.

\end{abstract}

\section{Introduction}
Infrared-bright galaxies have been found both locally ({\em e.g.} Arp
220 \& IRAS ultraluminous-infrared galaxies) and at high redshifts
($z\gtsim1$) in field and cluster-search surveys ({\em e.g.} Francis
{\em et al.} 1997; Spinrad {\em et al.}  1997).  The first {\em deep}
field infrared surveys by Elston {\em et al.}  (1988 \& 1989) and more
recent ones (Songaila {\em et al.} 1994; Glazebrook {\em et al.} 1995;
Moustakas {\em et al.} 1997) have all revealed small numbers of very
faint galaxies with dramatic optical$-$near-infrared colors.  Many of
these are also red in their optical$-$optical colors, and are
consistent with being passively evolving ellipticals at $z\sim1$.
However, there is also a curious population of faint ($I\gtsim24$)
galaxies that are infrared-bright ($I-K\gtsim4$), yet with relatively
{\em blue} optical colors ($V-I\ltsim2.5$; see Figure~9 of Moustakas
{\em et al.} 1997).  The colors of these ``faint red outlier
galaxies'' (\frogs) may derive from exceedingly old underlying stellar
populations, a dust-embedded starburst or AGN, or a combination
thereof.  In either case, the dramatic $I-K$ color strongly implies
that the $4000$\AA\ break lies between the $I$- and $K$-bands, which
implies that $1<z<4.3$.  Models strongly suggest that they are
actually at $1\ltsim z\ltsim3$.  This redshift range falls between
$z<1$ redshift-survey and the $z\gtsim3$ galaxies being routinely
discovered now (Steidel {\em et al.} 1996).  Our goal is to find what
connections \frogs\ and ``extremely red objects'' (\eros; $R-K>6$;
$I-K\sim6.5$) have between each other and with the observed galaxy
types at both high- and low-redshift ends.  Therefore, we have
undertaken an observational and modeling program to characterize the
nature of the \frogs, on whose progress we report here.

\section{Observations}
As the \frogs' colors and morphology are relatively nondescript in the
optical, identifying targets requires both deep optical and
near-infrared imaging.  Our original sample was drawn from deep
ground-based NTT (optical) and Keck (near-infrared) data (Moustakas
{\em et al.} 1997).  More recently, we have studied candidates in the
HDF (Figure~1) using public and archival data, and especially in the
deep ``Westphal'' pointing of the Groth Survey Strip (the ``dGSS'';
Figure~2) using the archival WFPC2 data as a starting point.  With
candidates identified in the latter field, we have obtained
high-resolution near-infrared imaging with NICMOS (F160W on NIC2) and
have (as yet unsuccessfully; see Table~1) attempted to obtain
redshifts using multislit spectroscopy with Keck/LRIS.

\begin{figure}[h]
\psfig{figure=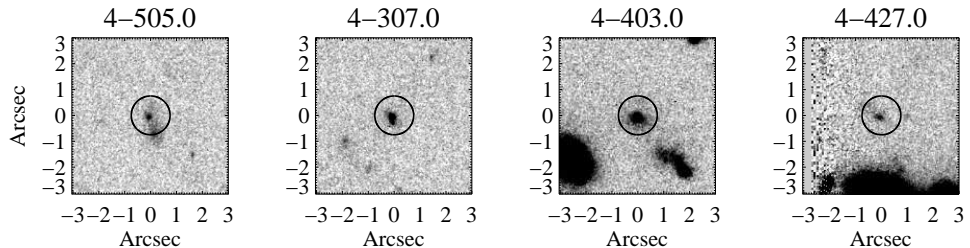}
\caption{{\small The four \frog\ candidates in the Hubble Deep Field
(F814W), identified using the optical WFPC2 photometry and the deep
$K$-band imaging of Hogg {\em et al.} (1997).  Their optical
morphologies are disturbed, suggesting that they are not relaxed,
uniformly old systems.  Their near-infrared (HST/NICMOS) morphologies,
and especially spectroscopy, are necessary follow-up observations to
help determine the nature of the underlying stellar populations and
their dust content.}}
\end{figure}

\begin{table}
\caption{\centerline{Summary of Observations in the {\bf {\em dGSS}}
Field (Moustakas {\em et al.}, in prep.)}}
\scriptsize
\begin{tabular}{llllrll}
\tableline
Date   & Telescope & Instrument & Setting & Exposure & Seeing & Remarks \\
\tableline
94 Mar & HST    & WFPC2  & F814W   &  23,100s & $\sim0.15''$ & Westphal GO-5109\\
94 Mar & HST    & WFPC2  & F606W   &  24,400s & $\sim0.14''$ & Westphal GO-5109\\
96 Apr & Keck   & NIRC   & $K (2.2\mu m)$     &  4,000s  & $\sim0.5''$ & Excellent Conditions\\
97 May & Keck   & LRISm  & 300~mm$^{-1}$  & $\sim 8^h$ & $\sim0.7''$ & Unphotometric\\
97 Sep & HST    & NICMOS & F160W   & 12,031s  & $\sim0.2''$ & Moustakas GO-7460\\
\tableline
\end{tabular}
\end{table}

\section{The Nature of the Faint Red Outlier Galaxies}

The HST-determined optical and near-infrared \frog\ morphologies
(Figures 1 \& 2) are compact ($\sim0.6''$) and asymmetric, suggesting
that they are not relaxed systems; and that the infrared ``excess'' is
likely primarily due to dust.  In that case, the nuclear region must
be very (infrared-) bright relative to the (blue-ish) ``disk,'' and
may be so either because of a nuclear starburst or an AGN (see {\em
e.g.}  Francis {\em et al.}  1996).  Spectroscopic redshifts and
features are necessary for further detailed study.

It is worth emphasizing that \frogs\ are selected {\em in the field},
and not in the vicinity of known quasars or AGN.  Most of the \eros\
are in such fields (Hu \& Ridgway 1994; Soifer {\em et al.} 1994;
Graham \& Dey 1996)\footnote{However, at least one survey, the CADIS
$K'$ Survey of Beckwith {\em et al.} (this conference) has produced
several \eros\ in untargetted fields.}.  Establishing the relation(s)
between the several flavors of infrared-excess galaxies will be very
important for understanding both the ages of high-redshift galaxies
({\em e.g.}  Spinrad {\em et al.}  1997; Dunlop {\em et al.} 1996),
the role dust plays as a function of redshift, and perhaps the
frequency of AGN activity in galaxies at high redshift.

\begin{figure}[h]
\psfig{figure=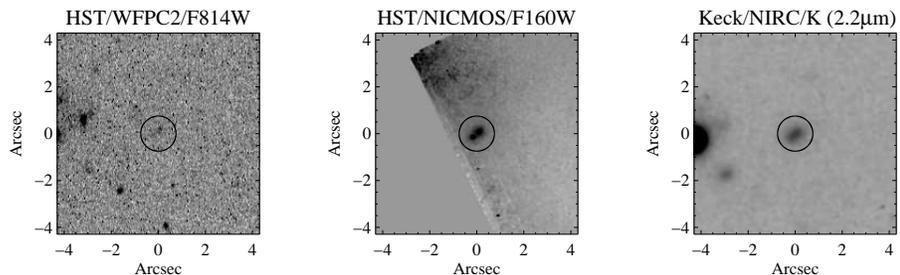,width=5.0in}
\caption{{\small A \frog\ candidate in the Westphal pointing of the
Groth Survey Strip. The optical F606W and F814W HST/WFPC2 data were
retrieved from the Archive (AR \#7524); the F160W HST/NICMOS image is
from recently completed observations under the Guest Observer program
\#7460 (Moustakas {\em et al.}); and the deep $K$-band ($2.2\mu m$)
image is a portion of a large, deep mosaic in the dGSS, acquired in
1996 April at the Keck~I Telescope with NIRC.  An (underway) study of
\frogs' optical \& near-infrared morphologies and surface-brightness
profiles will help distinguish between {\em age} and {\em dust} as the
causes of the infrared-excess light.}}
\end{figure}

\subsection{Motivation for Dusty \& Disky Morphologies}

The relative paucity of detections in Ly$\alpha$-protogalaxy searches
and mm observations of high-$z$ galaxies show that dust is very
important even at the earliest times.  Spheroid formation may be
triggered by mergers, which generate a central concentration of gas,
which in turn induces a central starburst.  This remains reddened by
the produced dust, even if the {\em gas} is driven out by winds and
shocks.  At a later time, a disk forms by gas infall.  This scenario
(see {\em e.g.} Wang \& Silk 1994) predicts that early on an
early-type galaxy may appear as a dusty, evolved nucleus, surrounded
by a young (blue) disk.  This composite produces the correct \frog\
colors, assuming $\sim1$~mag of extinction in the $K$-band in the
nuclear regions.

\section{Conclusions}

The \frogs\ are likely very dusty, $z>1.2$, systems, with a bright
nucleus which is powered either by a (low-level) AGN or a starburst.
They make up an appreciable fraction of faint infrared-selected
samples ($\ltsim$10\% of all galaxies at $K\sim20-22$), and so may
represent a significant epoch or stage in the history of galaxy
evolution.  They are more numerous (by $\sim10\times$) than \eros,
although their colors are less extreme.  Recently-obtained HST/NICMOS
observations yield near-infrared morphological information, which
combined with archival HST/WFPC2 images will provide clues for the
origin of the ``excess'' infrared flux.  Spectroscopic redshifts for
these very faint galaxies have so far proven elusive, even with a
dedicated effort with Keck.  This will be an ideal project to pursue
with next-generation ground- and space-based instrumentation,
particularly NIRSPEC on Keck and SIRTF from space.

\acknowledgments

We thank the Organizing Committee for a very productive meeting.  LAM
particularly thanks the STScI for travel and attendance support
through Program GO-7460.


\begin{references}

 \reference Dunlop, J. {\em et al.} 1996, Nature, 381, 581
 \reference Elston, R., Rieke, G.H, \& Rieke, M.J. 1988, \apj, 331, L77
 \reference Elston, R., Rieke, M.J, \& Rieke, G.H. 1989, \apj, 341, 80
 \reference Francis, P.J., Woodgate, B.E., \& Danks, A.C. 1997, \apj, 482, L25
 \reference Francis, P.J. {\em et al.} 1996, \apj, 457, 490
 \reference Glazebrook, K. {\em et al.} 1995, \mnras, 275, 169
 \reference Graham, J.R. \& Dey, A. 1996, \apj, 471, 720
 \reference Hogg, D.W. {\em et al.} 1997, \aj, 113, 474
 \reference Hu, E.M. \& Ridgway, S.E. 1994, \aj, 107, 1303
 \reference Moustakas, L.A. {\em et al.} 1997, \apj, 475, 445
 \reference Soifer, B.T. {\em et al.} 1994, \apj, 420, L1
 \reference Songaila, A., Cowie, L.L., Hu, E.M., \& Gardner, J.P. 1994, \apjs, 94, 461
 \reference Spinrad, H. {\em et al.} 1997, \apj, 484, 581
 \reference Steidel, C.C. {\em et al.} 1996, \apj, 462, L17
 \reference Wang, B., \& Silk, J. 1994, \apj, 427, 759

\end{references}
\end{document}